\documentclass[11pt,preprint]{aastex}

\slugcomment{To appear in Astrophysical Journal}
\shorttitle{Flat Cluster Cores and Fermionic Dark Matter}
\shortauthors{Nakajima and Morikawa}

\begin{document}

\title{An Interpretation of Flat Density Cores of Clusters
of Galaxies\\
 by Degeneracy Pressure of Fermionic Dark Matter: \\
A Case Study of Abell 1689}

\author{Tadashi Nakajima}
\affil{National Astronomical Observatory of Japan \\ 
Osawa 2-21-1, Mitaka, 181-8588, 
Japan}

\and 

\author{Masahiro Morikawa}
\affil{Department of Physics, Ochanomizu University \\
2-1-1 Otsuka, Bunkyo, Tokyo,112-8610, Japan
}

\begin{abstract}
Flat density cores have been obtained for
a limited number of clusters of galaxies by strong
gravitational lensing.
Using a phenomenological equation of state (EOS) describing
the full-to-partial degeneracy,
we integrate the equation of
hydrostatic equilibrium.
The EOS is based on an assumption that
the local kinetic energy of a classical particle induced
by the gravity 
dissolves the quantum statistical degeneracy.
The density profile is uniquely determined by four parameters,
the central density, $\rho(0)$, the properties of
a fermion,
namely, the mass, $m$, and 
statistical weight, $g$, and the ratio of the total matter density
and fermion density, $\delta$.
As a case study, we model the column density and 2D encircled
mass profiles of A1689, whose column density profile
has been observationally obtained by Broadhurst et al.,
using gravitational lensing.
The column density and 2D encircled profiles at the core, 
are reasonably reproduced for models with a limited range
of particle properties.
In the case that previously unknown fermions with spin 1/2 dominate
the dark matter, the acceptable particle mass range is between 2 and 4 eV.
In the case that the dark matter consists of the mixture of 
degenerate relic neutrinos and classical collisionless cold dark matter particles,
the mass range of neutrinos is between 1 and 2 eV, if the ratio of the two kinds
of dark matter particles is fixed to its cosmic value.
Both the pure fermionic dark matter models and neutrino-CDM-mixture models reproduce the observations equally well. 
 \end{abstract}


\section{Introduction}

In most of the cosmological studies, dark matter particles
are treated as collisionless particles. 
For instance, to describe the cosmological density evolution,
collisionless Boltzmann equation is adopted.
In cosmological N-body simulations, a group of 
collisionless dark matter particles, are represented by
a single collisionless particle in the computer from
the coarse-grained point of view.
In the treatment by the Boltzmann equation, in which
the distribution function is defined in a phase space,
the particle mass and statistical weight 
are explicitly dealt with, although the dark matter
particles are still regarded as classical particles without interaction
except for gravity.
In the coarse-grained view adopted by N-body simulations,
there is no particle information and only the global
mass density distribution is obtained.

However, by adopting the assumption that dark matter particles
are collisionless classical particles, we might have
lost basic physics in some cases.
At high number densities and relatively low temperature, low-mass elementary particles,
experience quantum statistical degeneracy due to indistinguishability of identical particles.
For instance, it used to be well known that the neutrino black body
in the early universe was partially degenerate \citep{Weinberg1962}.
Neutrinos are fermions and the radiation pressure of
the neutrino black body can be interpreted as the combination
degeneracy pressure and thermal pressure. As we discuss later in this paper,
massive relic neutrinos are likely to remain partially degenerate, after decoupling
and even after they become nonrelativistic under adiabatic expansion.

After nonlinear evolution of a high-density part of the universe, we might see a high concentration
of dark matter particles around the center of a cluster of galaxies. If dark matter particles are 
composed at least partially of light fermions (e.g. neutrinos), their degeneracy pressure
may be large enough to support the density structure near the center of the cluster against gravity.
A self-gravitating system supported by degeneracy pressure of fermions, such as
a white dwarf or a neutron star, is known to have a flat-top density profile. 
This is our motivation to explore the possibility that
recent results regarding the mass profiles of clusters of galaxies obtained
by gravitational lensing
\citep{Tyson98,Sand02,Sand04,TB05A,TB05B} might be explained by the degeneracy pressure
of light fermionic dark matter particles.

In this paper,
using a phenomenological equation of state (EOS) that describes
the physical conditions between fully degenerate fermionic gas
and the classical ideal gas, we integrate the equation of
hydrostatic equilibrium,
under the simple assumption that
the local kinetic energy of a classical particle is equal
to its gravitational energy determined by the 3D encircled mass.
Our model is expected to be valid only near the core
of a cluster where dynamical equilibrium is possibly achieved.
For pure fermions,
the volume density profile is uniquely determined by three parameters,
the central density, $\rho(0)$, and the properties of
dark matter particles,
namely, the mass, $m$, and 
statistical weight, $g$.
To compare our model with observations, we smoothly connect
our model volume density profile describing
the inner region to a volume density profile
derived from the observed column density profile by
assuming spherical symmetry at a radius
near the Einstein radius. In this way our model
column density profiles can be directly compared
with strong lensing results.
In reality, there are baryons and possibly 
other forms of dark matter such as massive cold dark
matter particles.
These general cases are dealt with by introducing
another input parameter, the ratio of
the total mass density and light fermion density,
$\delta$.

A1689 studied by \citet{TB05A,TB05B} is the best studied
cluster by both strong and weak lensing, for which column density
profile is obtained from the core to radii greater than 1 Mpc.
As a case study, we apply our modeling to this cluster
and constrain the possible combinations of particle properties.
As a natural candidate for light fermions, we consider the
case of massive neutrinos.

The paper is organized as follows. We first obtain 
the volume density of A1689 from the observed column density profile
of \citet{TB05B} and show that eV fermions can be degenerate
near the core of this cluster in \S2. Our modeling procedure
is described in \S3 and the properties of input fermions
are discussed in \S4, and then
the results are presented in \S5. 
The moderate degeneracy of unbound relic neutrinos and the plausibility
that they can fall into the cluster core, are discussed   
in \S6.

\section{Derivation of a Volume Density Profile
of A1689 from the Column Density
Profile of Broadhurst et al.}

Recently Broadhurst et al. \citep{TB05A,TB05B} 
reported a mass column density profile 
of the cluster of galaxies, 
A1689, obtained from gravitational lensing.
One of the important properties of the profile is that
it has a flat top.
We propose that this flat-top column density profile 
might be explained by the effects of degeneracy pressure of
fermionic dark matter. Here we analyze this proposal.

First we briefly introduce the main results of \citet{TB05A,TB05B}.
In their analysis, 1$^{\prime}$ corresponds to 129 kpc $h^{-1}$.
In \citet{TB05A}, the central 250 kpc $h^{-1}$ in radius
of multi-color HST/ACS images were analyzed.
The mass column density profile, $\Sigma(r)$,
is not expressed as a single power law
of radius.

The mass column density profile flattens toward the center with
a mean slope of \\
$d \log \Sigma / d\log r \approx -0.55$ 
within 
$r<$250 kpc $h^{-1}$. Inside the Einstein radius 
($\theta_E\approx 50^{\prime\prime}$), they obtained the slope
of $\approx -0.3$ from the ratio between 
$\theta_E$ and the radius of the radial critical curve,
$\theta_r\approx 17^{\prime\prime}$. 
They fit their results with an inner region of an NFW profile
\citep{NFW96}
with a relatively high concentration, $C_{vir} = 8.2$.

The mass column density, $\Sigma(r)$, is the integral
of the volume density, $\rho(r)$, 
along the line of sight over the entire cluster scale
of Mpc. In order to study the possibility of fermion degeneracy
near the center of the cluster,
we need information on the volume density, $\rho(r)$, instead of
the column density, $\Sigma(r)$.
\citet{TB05B} present the weak-lensing analysis of
the wide field data obtained by Subaru and obtained 
the column density profile at $r<2$ Mpc $h^{-1}$.
They fit the combined profile of HST/ACS and Subaru with
an NFW profile with a very high concentration,  $C_{vir} = 13.7$,
significantly larger than theoretically expected value of $C_{vir} \approx 4$.
They also fit the same observed column density profile with
a power law profile with a core. 
They give this result in terms of the angular radius dependence 
of the convergence, $\kappa$, as

\begin{equation}
\kappa \propto (\theta + \theta_C)^{-n}.
\label{kappa}
\end{equation}

$\theta_C = 1.65^{\prime}$ and $n = 3.16$ give the best fit although
$\theta_C$ and $n = 3.16$ are mutually dependent and a finite range
of the combination ($\theta_C$,$n$) gives equally good fits.
In terms of $\chi^2$ and the degrees of freedom, this
core power law profile fits the observation better than
the best-fit NFW profile and we use this profile for further discussion.
Although \citet{TB05B} do not claim so explicitly, the two facts that
the best-fit NFW profile shows a much higher concentration than
the value predicted by the CDM cosmology and the phenomenological
profile, eq.(\ref{kappa}), fits better than the best-fit NFW profile,
may indicate some contradiction to the CDM cosmology.

We start
our analysis from 
this core power-law profile, eq(\ref{kappa}), for further discussion.
We convert (\ref{kappa}) to a column density profile, $\Sigma(r)$,
in physical units of length and mass using the relations,
$\kappa = \Sigma/\Sigma_{crit}$, $\Sigma_{crit} \approx$ 0.95
${\rm g}\cdot{\rm cm}^{-2}$, and the normalization of 2D encircled
mass inside
the Einstein radius, $r_E$,
$\int_0^{r_E} \Sigma(r) 2\pi r dr = \Sigma_{crit}\cdot \pi r_E^2$.
The result is expressed as

\begin{equation}
\Sigma(r) = 25.2 \cdot \left(r/r_E + 2.2\right)^{-3.16}, 
\hskip 1cm ({\rm g}\cdot{\rm cm}^{-2})
\label{cpl}
\end{equation}

where $r_E = 97$ kpc $h^{-1}$
corresponds to $\theta_E=45^{\prime\prime}$,
the value used in \citet{TB05B}. 
The 2D encircled mass, $M_2(r) = \int \Sigma(r) 2 \pi r dr$, is
analytically obtained and  $M_2(r) = 1.3\times 10^{14} h^{-2} M_\odot$ and
$1.1\times 10^{15}h^{-2} M_\odot$ respectively for $r = r_E$ and $r=\infty$.
Therefore a high concentration of the mass is expected on
the scale of $r_E$. By assuming
spherical symmetry, we wish to obtain the volume density $\rho(r)$ by
solving 

\begin{equation}
\Sigma(x) = \int \rho(\sqrt{x^2+z^2}) dz.
\end{equation}

Instead using a standard method like
the Abel transform, we assumed another power-law profile with a core radius for $\rho(r)$
and obtained the best fit parameters, to see the changes of the core radius
and the power law index. From now on, we fix $h=0.73$ \citep{WMAP}, so that
direct comparison between observations and models can be made. 
The range of integration
in $z$ is from $-$1.4Mpc to $+$1.4Mpc, 
which correspond to the region with good weak lensing signals,
since we need solid numbers on the scale of cluster core. 
The result is

\begin{equation}
\rho(r) = 5.1\times10^{-24} \left(r/r_E + 1.077\right)^{-3.41}.
\hskip 1cm ({\rm g}\cdot{\rm cm}^{-3})
\label{3density}
\end{equation}

The best fit parameters somewhat depends on the range
of integration. For instance, if we use the observed column density
profile eq.(\ref{cpl}) out to $\pm$2.8 Mpc, regardless of the  
strength of weak lensing signals, the  
core radius in units of $r_E$, and power-law index, change to
1.25, and -3.71, respectively. 
Note that the power-law index of our choice, -3.41, is
closer to that of an NFW profile of -3, than -3.71.
Since our interest is in the inner region as we show later, and 
since the contribution of the outer region to the total mass
is small, our choice should be justified.

\section{Modeling Procedure}

\subsection{Degeneracy of eV-Mass Fermions}

Before proceeding to modeling of mass profiles,
we first show that at the center of A1689 with volume densities
of order of $10^{-24}$ (g$\cdot$cm$^{-3}$), nonrelativistic
eV-mass fermions can become degenerate.
Since a mass of 1 eV corresponds to $1.8\times10^{-33}$ g, the number density, 
$N/V \approx 10^{11}$ cm$^{-3}$, and the mean inter-particle spacing is,
$(N/V)^{-1/3} \approx 2\times 10^{-4}$ cm. On the other hand,
the de Broglie wavelength for a 1 eV particle with a relative velocity $v$ is,
$h/\mu_0 v = h/(\mu_0c)(c/v)=\lambda_{Compton}\cdot(c/v) = 1.2\times10^{-4}(c/v)$ cm. 
Therefore for nonrelativistic particles with $v \ll c$, the condition 
for high degeneracy, $(N/V)^{-1/3} \ll \lambda_{({\rm de \hskip 5pt Broglie})}$, 
is satisfied. We first formulate the modeling procedure of matter distribution
for the case that the entire matter consists purely of fermionic dark matter and
then, modify the formulation for the case that the fractional contribution
of fermionic dark matter density to the total matter density is constant.

\subsection{Phenomenological Equation of State}

First we provide our justification for introducing an equation of state and
assuming hydrostatic equilibrium
for the mixture of degenerate fermions and non-degenerate classical
collisionless particles. 
A self-gravitating system composed purely of classical collisionless particles
such as cold dark matter particles may be thermodynamically anomalous \citep{Lynden-Bell}
and the equation of state may be poorly defined. However,
the effect of fermion degeneracy or introduction of repulsion due to
Pauli's exclusion principle is to make the mixture of
degenerate fermions and classical collisionless particles a thermodynamically
normal system and an analysis based on hydrostatic equilibrium 
valid.

To deal with the general situations in which particle temperature is
finite and degeneracy is partial, we need to know the equation of state (EOS),
and have to determine the temperature profile along with the density
profile. We adopt two major assumptions that simplify our
analysis of fermionic dark matter distribution.
First, we assume that the EOS, or the pressure law, has the following form,

\begin{equation}
 p = p_D + n k_B T,
\label{peos}
\end{equation}

where $p_D$ is the zero temperature degeneracy pressure given by

\begin{equation}
p_D = \frac{1}{5} \left(\frac{6\pi^2}{g}\right)^{2/3}
\frac{\hbar^2}{m} n^{5/3},
\label{pd}
\end{equation}

$n$ is the number density and $n k_B T$ is 
the thermal pressure of the classical ideal gas.
Eq.(\ref{peos}) can be rewritten as,

\begin{equation}
p = p_D \left(1 + \frac{5}{2} \frac{k_B T}{E_f} \right),
\label{peos2}
\end{equation}

where $E_f$ is the Fermi level given by

\begin{equation}
E_f = \left(\frac{6\pi^2}{g}\right)^{2/3}
\frac{\hbar^2}{2m} n^{2/3},
\label{ef}
\end{equation}

and ${E_f}/(k_B T)$ corresponds to the degree of degeneracy. The formulae for degeneracy
pressure and Fermi level, eqs.(\ref{pd},\ref{ef}), are derived in a textbook of statistical physics \citep{Landau}. 
This EOS, eq(\ref{peos}), has been used, for
example, in an analytic model of the brown dwarf
interior where electrons are expected to be
partially degenerate \citep{Burrows93}. The approximation
used in this EOS may be crude for intermediate degeneracy.

The second assumption is with regard to classical kinetic energy.
Collisionless particles 
in general may not be in local thermodynamic equilibrium at
a given radius, $r$. As we have mentioned before,
the classical thermal pressure term in eq(\ref{peos})
for collisionless particles is actually the pressure equivalent,
$m n \sigma^2$, where $\sigma$ is the velocity dispersion 
in the static Jeans equation with an isotropic velocity dispersion \citep{Binney}. 
Physically quantum statistical degeneracy will be dissolved by
classical kinetic energy,
or velocity dispersion, $\sigma$, as $\sigma$ increases.
Here we further assume that the particle kinetic energy at $r$, 
is in balance with the gravitational energy induced by the 3D encircled mass
interior to $r$. From this assumption, we obtain,

\begin{equation}
\frac{3}{2} k_B T = \frac{G m M(r)}{r},
\label{ASSUMPTION2}
\end{equation}

where 

\begin{equation}
M(r) = \int_0^r 4 \pi \rho(r) r^2 dr.
\label{3DMASS}
\end{equation}

From eqs.(\ref{peos},\ref{ASSUMPTION2}), we obtain the expression for 
the pressure,

\begin{equation}
p =  \frac{1}{5} \left(\frac{6\pi^2}{g}\right)^{2/3}
\frac{\hbar^2}{m} n^{5/3} + \frac{2}{3}\frac{G m n M(r)}{r}.
\label{plaw}
\end{equation}

\subsection{Equation of Dynamical Equilibrium}

For ordinary gas, the equation of hydrostatic equilibrium (EHSE) describes
the dynamical equilibrium. For collisionless particles, its counterpart
is the static Jeans equation with an isotropic velocity dispersion \citep{Binney}.
In the pressure law, eq(\ref{plaw}), the first term is the gas pressure of
degenerate fermions, while the second term is the pressure equivalent
of collisionless non-degenerate particles. Since, the EHSE and isotropic
Jeans equation exactly have the same mathematical form, eq(\ref{plaw}) can
be inserted to the EHSE,

\begin{equation}
\frac{dp}{dr} = -\rho(r) \frac{G M(r)}{r^2}.
\label{HSE}
\end{equation}

Below we give a brief description of how to integrate eq(\ref{HSE}).
First, we define $N(r)$ by

\begin{equation}
N(r) = \int_0^r n(r) r^2 dr,
\label{N}
\end{equation}

or in its differential form,

\begin{equation}
\frac{dN(r)}{dr} = n(r) r^2.
\label{dNdr}
\end{equation}

$N(r)$ is the 3D encircled particle number divided by $4 \pi$ and satisfies 
a boundary condition that $N(0)=0$.

Eq(\ref{HSE}) is reduced to 

\begin{equation}
\frac{dn(r)}{dr} = - 
\frac{\left[n(r) N(r) + 2 r^3 n(r)^2\right]}{\left[A r^2 n(r)^{2/3} + 2 r N(r)\right]},
\label{dndr}
\end{equation}

where 

\begin{equation}
A = \frac{1}{4\pi}\left(\frac{6\pi^2}{g}\right)^{2/3}
   \frac{\hbar^2}{G m^3},
\end{equation}

which has the dimension of length. Even with double precision, 
the dynamic range is too large for cgs units, to deal with
cluster scale quantities. We adopt kpc as the units of length.
In this case, $A = 4.36 \times 10^{28}$.
  
Parameters that determine a solution are 
the central number density $n(0)=\rho(0)/m$, 
which is positive and finite,
and particle properties, $m$ and $g$.
Integration of eqs.(\ref{dndr},\ref{dNdr}) is performed simultaneously,
using the fourth-order Runge-Kutta method. 

So far we have dealt with the case that the mass density is
purely due to fermionic dark matter. 
In reality, light fermions may account for a limited fraction of entire dark matter.
Moreover, the 
baryon-to-dark-matter ratio is about 1:5 on average for the entire universe,
and we have to take into account the baryon contribution.

Since we have no reliable information on
the radial dependence of 
$\delta = \rho({\rm total})/\rho({\rm fermion})$ 
near the cluster center,
we assume a constant $\delta$. 
By incorporating $\delta$ into eq(\ref{HSE}),
eq.(\ref{dndr}) is modified as

\begin{equation}
\frac{dn(r)}{dr} = - 
\frac{\left[\delta^2 n(r) N(r) + 2 \delta^2 r^3 n(r)^2 \right]}{\left[A r^2 n(r)^{2/3} + 2 \delta^2 r N(r)\right]}.
\label{dndr2}
\end{equation}

Later we also consider the case that $\delta$ depends on $r$ due to the contribution of a massive galaxy
placed at the center to see how much model parameters are affected by the presence of a galaxy.

\subsection{Applicability of the Phenomenological Equation of State}

Since the asymptotic form of the 
phenomenological EOS is that of classical ideal gas,
a model profile approaches to an isothermal profile or
$\rho(r) \propto r^{-2}$ at large $r$, which does not approximate the behavior
of the observed core-power-law 
profile at large $r$ 
($\rho(r) \propto r^{-3.41}$). 
The same is true for an NFW profile \citep{NFW97} for which
$\rho(r) \propto r^{-3}$ at large $r$.
So the validity of the model
is limited to the core of the cluster for which $r \sim r_E$.
Therefore we cannot compare our models directly with the column density
profile obtained by \citet{TB05B}, since we cannot integrate
a model volume density profile along an entire line of sight. 
On the other hand, the most robust observable obtained from
strong gravitational lensing is the 2D encircled mass
within the Einstein radius, $M_2(r_E)$, which gives the
best constraint on the model.
The second best observable is the power-law slope
of the column density profile between the radial
and tangential critical curves. These two quantities
are taken into account in deriving the observed density
profile, but again should be taken into account
in the comparison with the model as well.
In any case, it is best to compare our models with
the observed column density profile out to $r \sim r_E$.

\subsection{Smoothly Connecting an Inner Model Profile with the Outer Observed Profile}

We construct a hybrid column density profile by smoothly connecting
a model volume density profile with the observed volume density
profile at a transition radius $r_t$.
The actual procedure is that the radius, $r_t$, is searched for at
which both $\rho(r_t)$ and $d\rho/dr(r_t)$ are connected.
This condition is physically justified below.
We derive a model profile under the assumption of the hydrostatic equilibrium
(HSE). In order for HSE to hold also at $r_t$, the pressure
gradient, $dp/dr$ needs to be connected as well as $\rho$.
If the EOS has the form that 
the pressure $p$ is given as an analytic function of $\rho$ as

\begin{equation}
p = f(\rho),
\label{normalEOS}
\end{equation}

its derivative is 

\begin{equation}
\frac{dp}{dr} = \frac{df(\rho)}{d\rho} \frac{d\rho}{dr}.
\end{equation} 

Therefore
the continuity of $dp/dr$ is equivalent to that of  $d\rho/dr$ under
the condition that $\rho$ is continuous.
Since the 3D encircled mass is the volume integral of $\rho(r)$,
the EHSE becomes a differential equation of second order,
when the pressure is given as eq.(\ref{normalEOS}).
This is the mathematical justification for both $\rho$ and $d\rho/dr$
to be connected. Our phenomenological EOS is of course analytic.

The observed profile is given as $\rho(r)$, but
the validity of EHSE is not guaranteed for entire $r$ of the cluster,
since the behavior of $p(r)$ or that of the velocity
dispersion $\sigma(r)$ is unknown. However in
the inner region of our interest, the EHSE is
expected to hold, as long as our model describes
the actual physical situation.

The computation in practice is implemented as follows.

\noindent
1. Give ($m,g,\delta$) as input parameters.

\noindent
2. Set the initial transition radius $r_t$ to 0.5$r_E$ = 66.4 kpc.
Since the location of the radial critical curve at 0.38$r_E$,
there is not strong observational constraint  inside 0.5$r_E$.
In terms of the observed 2D encircled mass, $M_2(0.5r_E)$,
is only 34\% of the robust observable, $M_2(r_E)$.

\noindent
3. Adjust $\rho(0)$ iteratively so that the agreement of the  
model and observed $\rho$s at $r_t$ is good to 5\%.

\noindent
4. Increase $r_t$ gradually until the matching of the model and observed $d\rho/dr$s 
becomes better than 5\%, while $\rho(0)$ is 
continually adjusted so that the connection of $\rho(r_t)$ is maintained.

\noindent
5. If the connection is found, the model and observed 2D encircled
masses within $r_E$, $M_2(r_E)$s are compared. 
When the agreement of the two encircled masses is found to be good to 2\%,
we accept this solution.

\section{Fermion Candidates, Particle Properties, and Mass Fraction of Fermions}

\subsection{Particle Mass Range of Interest}

Since the size of a self-gravitating degenerate star depends 
strongly on the particle mass as $m^{-2}$ \citep{OV39}, 
the allowed range of the particle mass is restricted to the order
of eV, if the degeneracy pressure is significant only near the flat-top 
cluster core, whose size is order of 100 kpc. This order estimation later turns out to be
correct, since solutions were found only for a narrow mass range
near 1 eV. Below we examine the possible ranges of statistical weight, $g$,
of fermions, and the fraction of the fermion density to the entire dark matter density.

\subsection{Dark Matter Dominated by  Unknown Weakly Interacting Fermions}

From SUSY, each known boson is expected to have a fermionic superpartner.
SUSY also predicts the stability of a lightest supersymmetric particle,
probably the photino. However, the
lack of experimental detection at energy scales less than 
10 GeV, indicates that photinos are massive and are more of 
a candidate for cold dark matter \citep{Peacock}.

Since there is no candidate predicted to exist theoretically,
we simply consider the case for a particle with spin 1/2
and assume that entire dark matter consists of these unknown fermions.
From the spin degrees of freedom, there are two possibilities, $g=1$ (Majorana fermion),
or $g=2$ (Dirac fermion). Another possibility is that statistically independent
particles and anti-particles equally contribute to the number density.
This gives effective $g=2$ (Majorana fermion) and $g=4$ (Dirac fermion).

\subsection{Massive Neutrinos}

The most natural candidate for which its presence
is well known is massive neutrinos. Due to the small mass differences
($\Delta m \le 0.05$ eV) among the three species, they
must have a similar mass (degenerate hierarchy),
if the approximate particle mass is as  large as 1 eV.
Neutrinos and anti-neutrinos are considered as distinguishable particles
and the mean number densities of relics particles are the same.
It will be natural to assume that the same ratio is maintained in the cluster core,
since there is no physical cause for segregation.
In terms of the effective statistical weight for the three neutrino species, 
$g = 6$ both for Majorana and Dirac neutrinos as long as we consider 
light relic neutrinos. Although a Dirac particle has two helicity states,
only right-handed particles were populated in the thermal equilibrium of
the early universe and
the current number density of left-handed particles are expected to be negligible
\citep{Lesgourgues}.

The current number density of each neutrino species (particles and anti-particles)
regardless of Majorana or Dirac neutrinos, is estimated to be 112.6 cm$^{-3}$.
If we adopt the cosmological parameters from three-year WMAP observations \citep{WMAP}, 
$h = 0.73$, $\Omega_\Lambda = 0.762$, $\Omega_{\rm DM} = 0.196$,
and $\Omega_{\rm B} = 0.042$, where DM and B stand for
dark matter and baryons respectively.
The mean mass density of the three species of neutrinos with the approximate mass, $m$ eV,
is $6.018\times 10^{-31} m$ g cm$^{-3}$, while the present critical density of
the universe 
is $1.0012\times 10^{-29}$ g cm$^{-3}$. 
Then, the neutrino mass fraction to the critical density, $\Omega_\nu$,
is given by,

\begin{equation}
\Omega_\nu \approx 0.060 m.
\label{omeganu}
\end{equation}

Near the cluster center, $R = \rho({\rm B})/\rho({\rm DM})$,
can be higher than the average value due to the dissipative collapse of baryons. 
Since $R$ at the cluster center is unknown, we adopt
the cosmic average, $R=0.2$, and to investigate how the change of $R$ 
affects model parameters,  we also model the case for $R=0.5$.
The spatial distribution of the cold dark matter
and light neutrinos may also be different, if neutrino distribution
is controlled by degeneracy pressure, while CDM particles behave as
classical collisionless particles. This effect is neglected for our
initial attempt and the $\rho({\rm DM})/\rho({\rm neutrino})$ is
fixed to the cosmic value, $\Omega_{\rm DM}/\Omega_\nu$.
With these assumptions,
the total mass density to neutrino mass density, $\delta$,
is related to $R$ as,

\begin{equation}
\delta = (1+R)\frac{\Omega_{\rm DM}}{\Omega_\nu} = 
(1+R)\frac{0.196}{0.060m}=(1+R)\frac{3.26}{m}.
\end{equation}

\subsection{Effect of a Massive Galaxy}

Although the total mass of galaxies in the cluster 
will be a small fraction of the total mass of
the cluster, a massive galaxy can affect
the local mass density significantly.
To study the influence of a massive galaxy
on model parameters, we add
a toy galaxy placed at the center to
the particle density and solve EHSE, and compare the
result with that without this hypothetical galaxy.

For simplicity, the density profile of this galaxy
is assumed to be given by a 3D Gaussian as, 

\begin{equation}
\rho_g(r) = \rho_{g0} \exp \left(-\frac{r^2}{2r_g^2}\right),
\end{equation}

where 

\begin{equation}
\rho_{g0} = \sqrt{\frac{2}{\pi}} \frac{M_g}{4\pi r_g^3},
\end{equation}

where $M_g$ is the mass of the galaxy and $r_g$ is the Gaussian scale length.
In modeling, $\delta$ is no longer a constant and depends on $r$ as,

\begin{equation}
\delta(r) = \frac{\rho_g(r)}{\rho({\rm fermion})} + \delta({\rm particle ~ only}).
\end{equation}

\section{Results}

\subsection{Dark Matter Dominated by a Single Species of Fermions}

\begin{figure}[tbp]
\includegraphics[angle=-90]{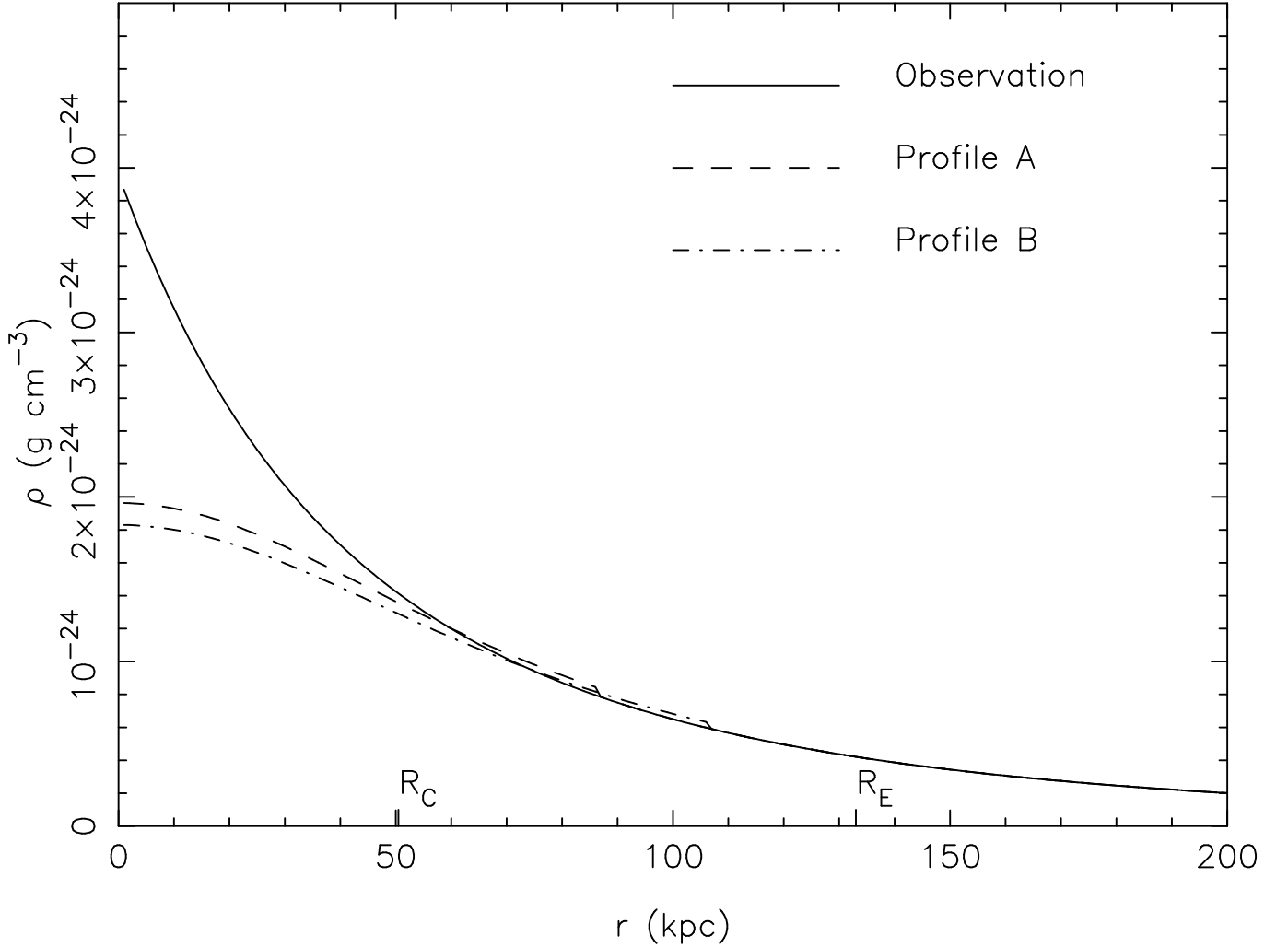}
\caption{Volume density profiles for general fermionic dark matter.
Profile A corresponds to
the highest density model and profile B is the lowest density model and other model profiles
lie between the two extremes (Profile I). The correspondence between 
a model profile and a solution is given in Table \ref{fermion.table}. 
The model profiles reproduces the observed profile well
at $r>50$ kpc. $r_c$ and $r_E$  stand for the radius of
radial tangential critical curve and the Einstein radius respectively.
\label{fermion.rho}}
\end{figure}

\begin{figure}[tbp]
\includegraphics[angle=-90]{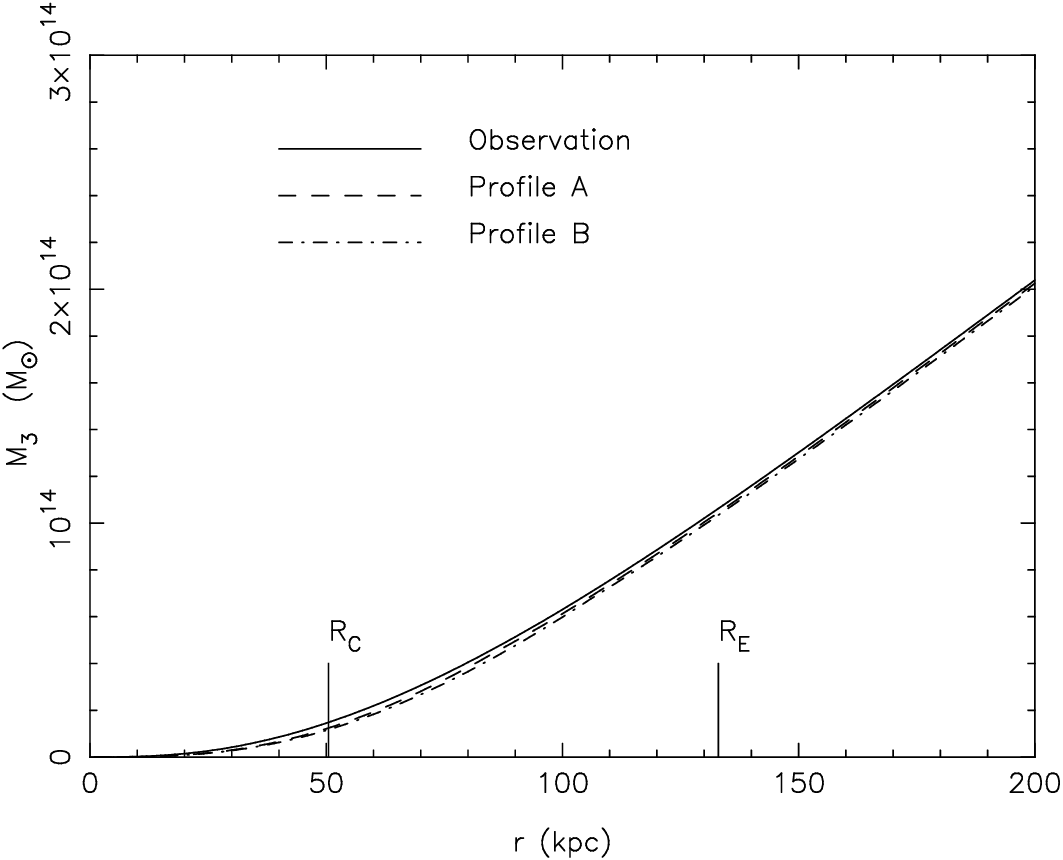}
\caption{3D encircled mass profiles for general fermionic dark matter.  The observed profile and
profiles A and B are indistinguishable. \label{fermion.M3}}
\end{figure}

\begin{figure}[tbp]
\includegraphics[angle=-90]{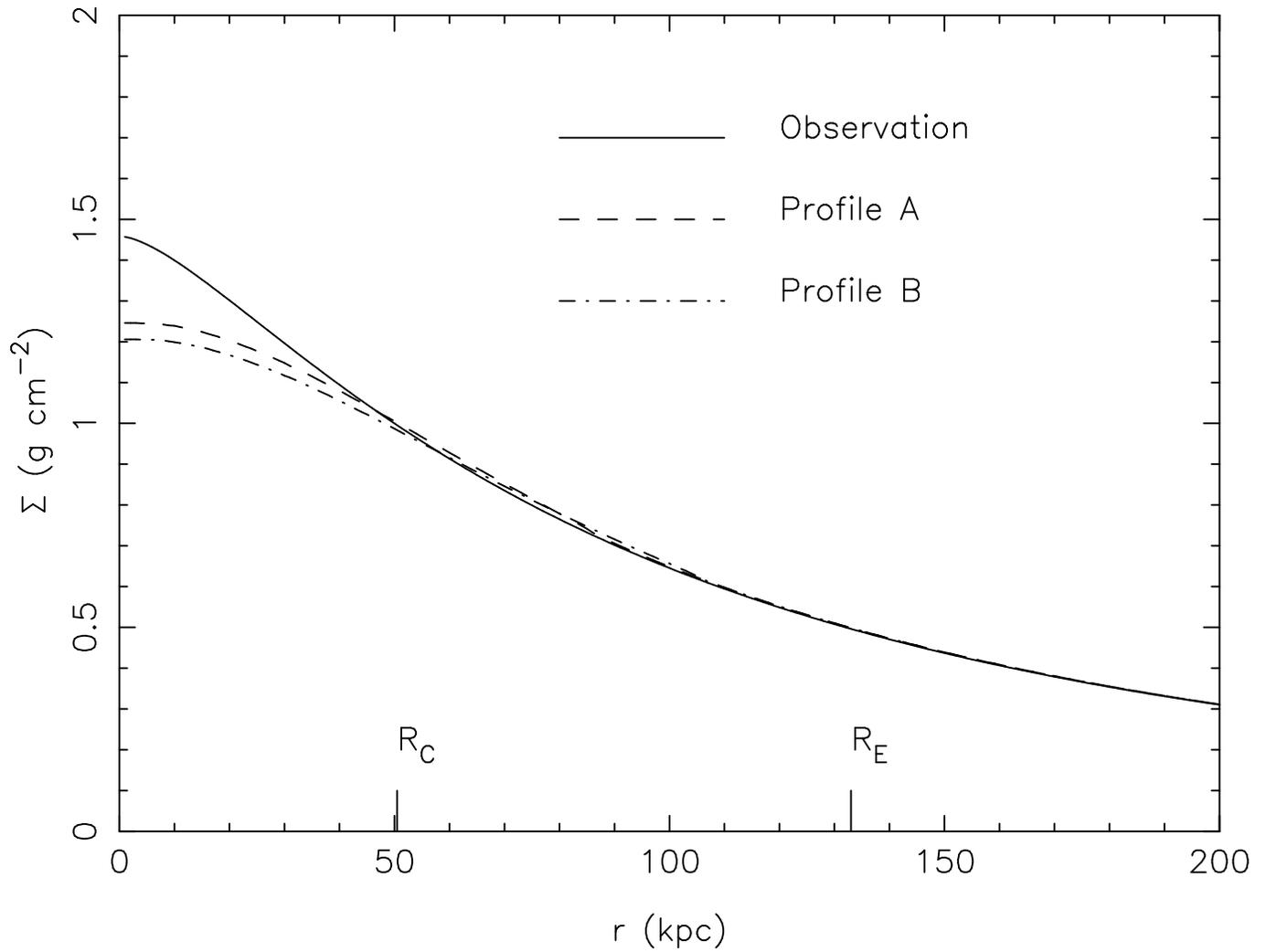}
\caption{Column density profiles for general fermionic dark matter.  
The model profiles reproduces the observed profile well 
at $r>50$ kpc.\label{fermion.Sigma}}
\end{figure}

\begin{figure}[tbp]
\includegraphics[angle=-90]{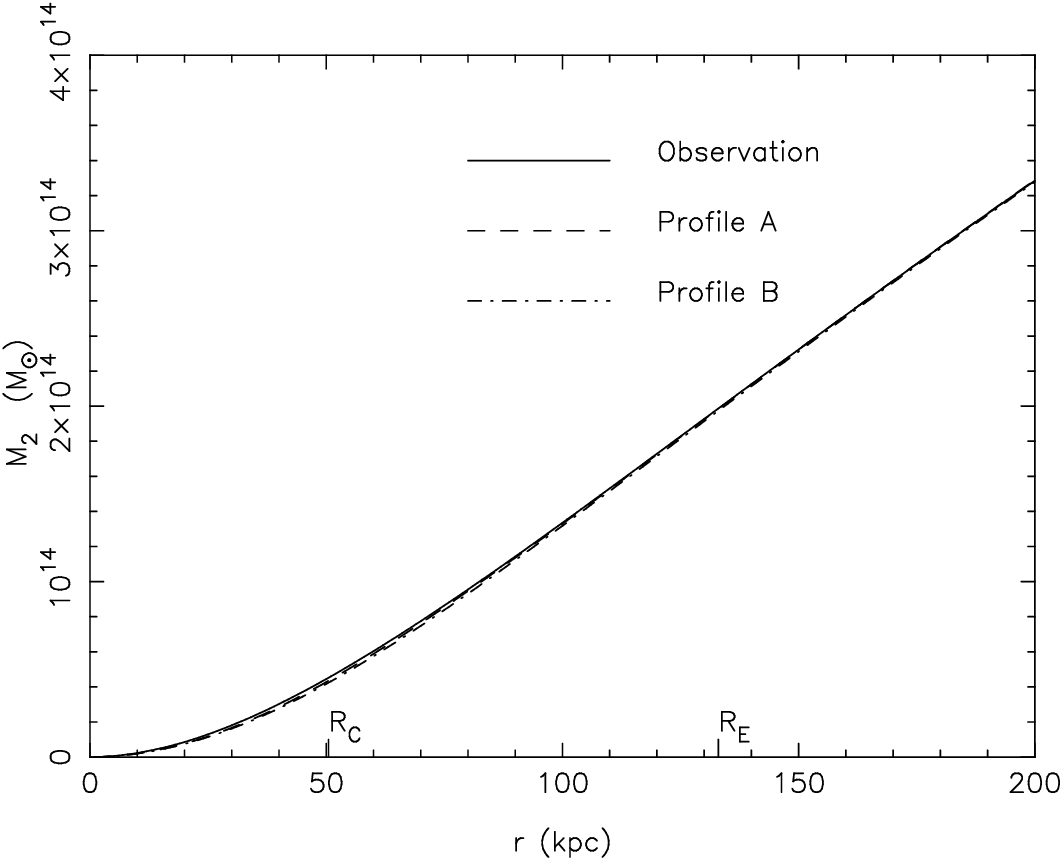}
\caption{2D encircled mass profiles for general fermionic dark matter.  The observed profile and
profiles A and B are indistinguishable. \label{fermion.M2}}
\end{figure}

Successful solutions are listed in
Table \ref{fermion.table}. 
The general trend is that for the same $g$, the smaller the particle mass, $m$, the larger the
transition radius, $r_t$. If we measure the significance of a model by the radial
coverage, which also corresponds the largest
encircled mass coverage,  the smallest $m$ solution is the most important for a given $g$.
For the smallest mass solutions, the range of particle mass is $m = 3.9 \sim 2.7$ eV for $g=1 \sim 4$
for the cosmic baryon-DM ratio of 0.2. A larger baryon fraction reduces the particle mass.
For the smallest $m$ solutions,
resultant volume- and column density profiles are almost identical as shown in 
Figs.\ref{fermion.rho} and \ref{fermion.Sigma}. There are some deficiencies
of mass densities inside the location of the radial critical curve, $r_c$.
However, in 3D and 2D encircled mass profiles
shown in Figs.\ref{fermion.M3} and \ref{fermion.M2}, these mass deficiencies,
which amount to at most 2\% of the 3D encircled mass 
inside $r_E$ ($2 \times 10^{12} M_\odot$), are not noticeable.

\subsection{Dark Matter Consisting of Massive Neutrinos and Cold Dark Matter}

\begin{figure}[tbp]
\includegraphics[angle=-90]{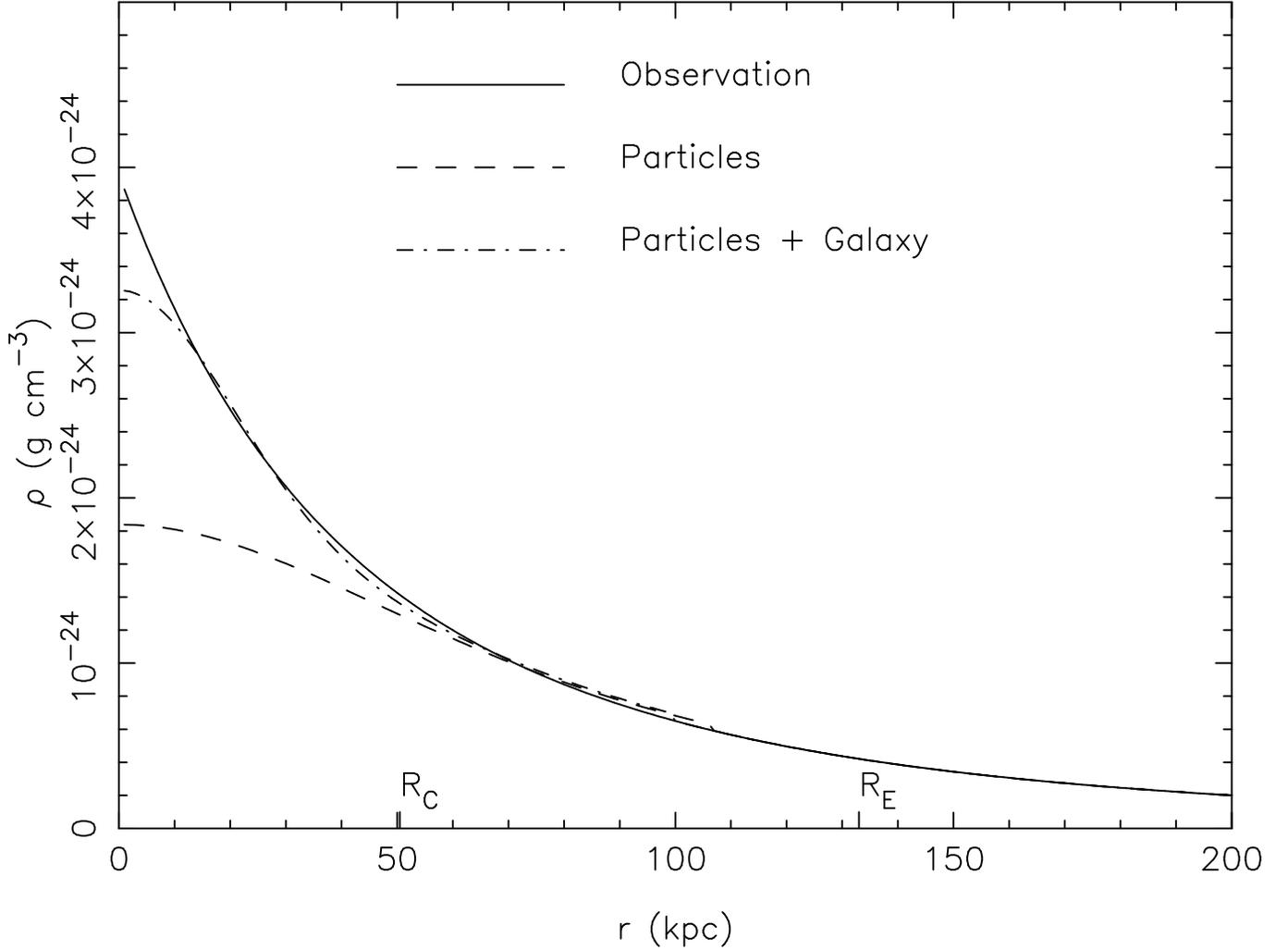}
\caption{Volume density profiles for massive neutrino models.
The solutions for neutrinos listed in Table \ref{neutrino.table}
have almost identical mass profiles. 
The observed profile is compared with the representative neutrino
solution (Particles = neutrinos + CDM particles + baryons) and
a solution with a toy galaxy (Particle + galaxy) placed at the center.
The mass of the toy galaxy is $2\times10^{12}M_\odot$,
2\% of the 3D encircled mass within the Einstein radius. 
The model without the galaxy reproduces the observation well except for the
central region, while the toy galaxy remedies the discrepancy.
It should be noted that the contribution of galaxies is included in
the observed density profile of \citet{TB05B}.
\label{neu.rho}}
\end{figure}

\begin{figure}[tbp]
\includegraphics[angle=-90]{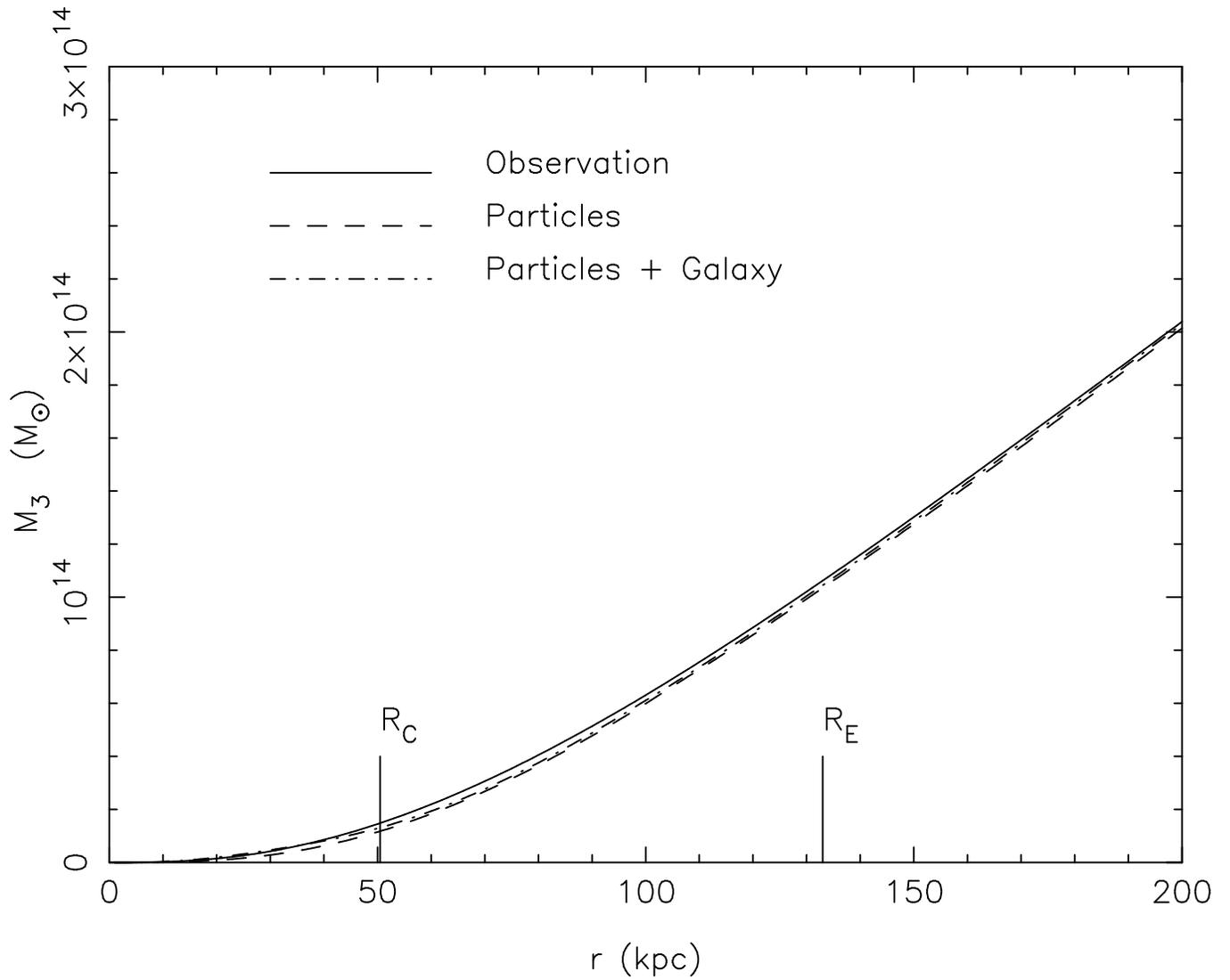}
\caption{3D encircled mass profiles for massive neutrinos models.
All three profiles are indistinguishable and the effect
of the central galaxy is not noticeable. \label{neu.M3}
}
\end{figure}

\begin{figure}[tbp]
\includegraphics[angle=-90]{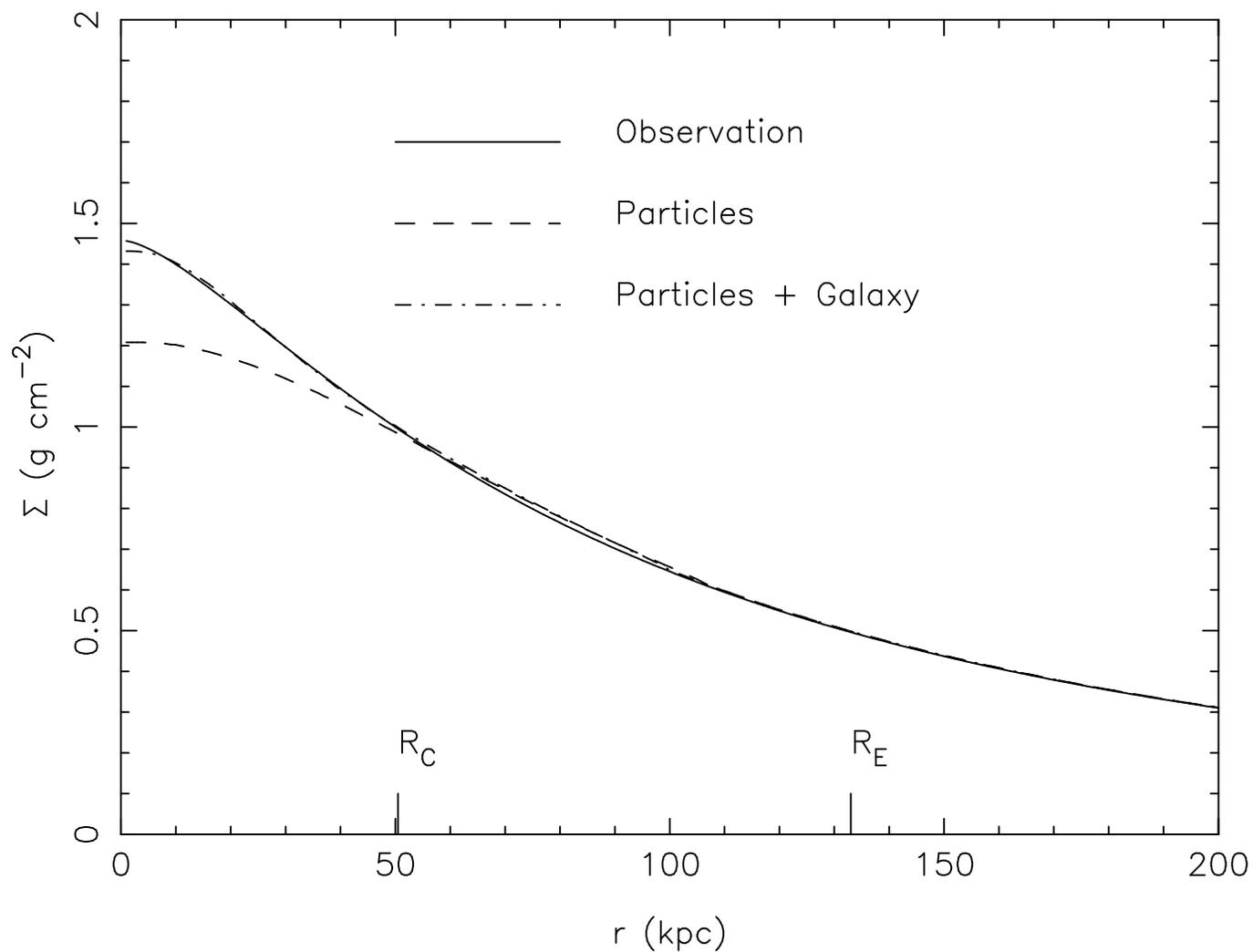}
\caption{Column density profiles for massive neutrinos.
Apart from the central region where the encircled mass is small,
The observed profile and the representative neutrino model profile
agree well. Again the toy galaxy fixes the discrepancy. \label{neu.Sigma}
}
\end{figure}

\begin{figure}[tbp]
\includegraphics[angle=-90]{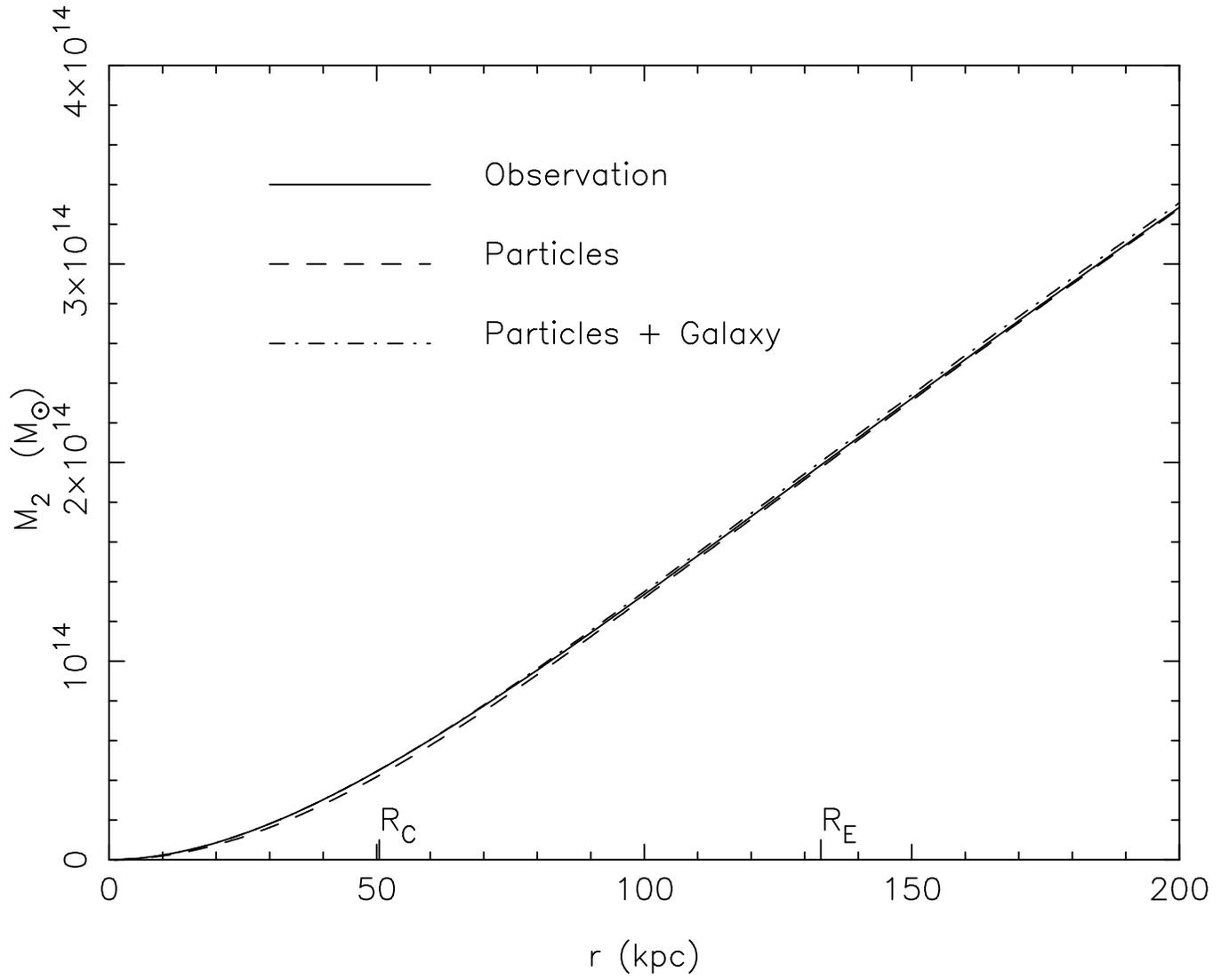}
\caption{2D encircled mass profiles for massive neutrinos.
All three profiles are indistinguishable. \label{neu.M2}
}
\end{figure}

The maximum allowed particle mass, $m$, is obtained by equating $\Omega_\nu$ and $\Omega_{\rm DM}$
as 3.27 eV. As can be seen in Table \ref{neutrino.table}, successful solutions have smaller particle
masses. Therefore we are dealing with the cases that dark matter consists of neutrinos and another form
of particles, possibly, nondegenerate cold dark matter particles.
The volume and column density profiles are given in Figs.\ref{neu.rho} and \ref{neu.Sigma} respectively.
Again the model profiles are always lower than the observed profile near the center.
However, the discrepancy between the models and observation is negligible in the linear plots of
the encircled masses shown in Figs.\ref{neu.M3} and \ref{neu.M2}.
Although $m$ depends on the baryon-DM ratio, $R$, the resultant profiles are essentially
identical and they are not distinguishable from the general fermion cases.
For $R=0.2$ and 0.5, the smallest mass solutions are for $m = 1.6$ and 1.1 eV respectively.

\subsection{Effect of a Toy Massive Galaxy on Density Profiles} 

\begin{figure}[tbp]
\includegraphics[angle=-90]{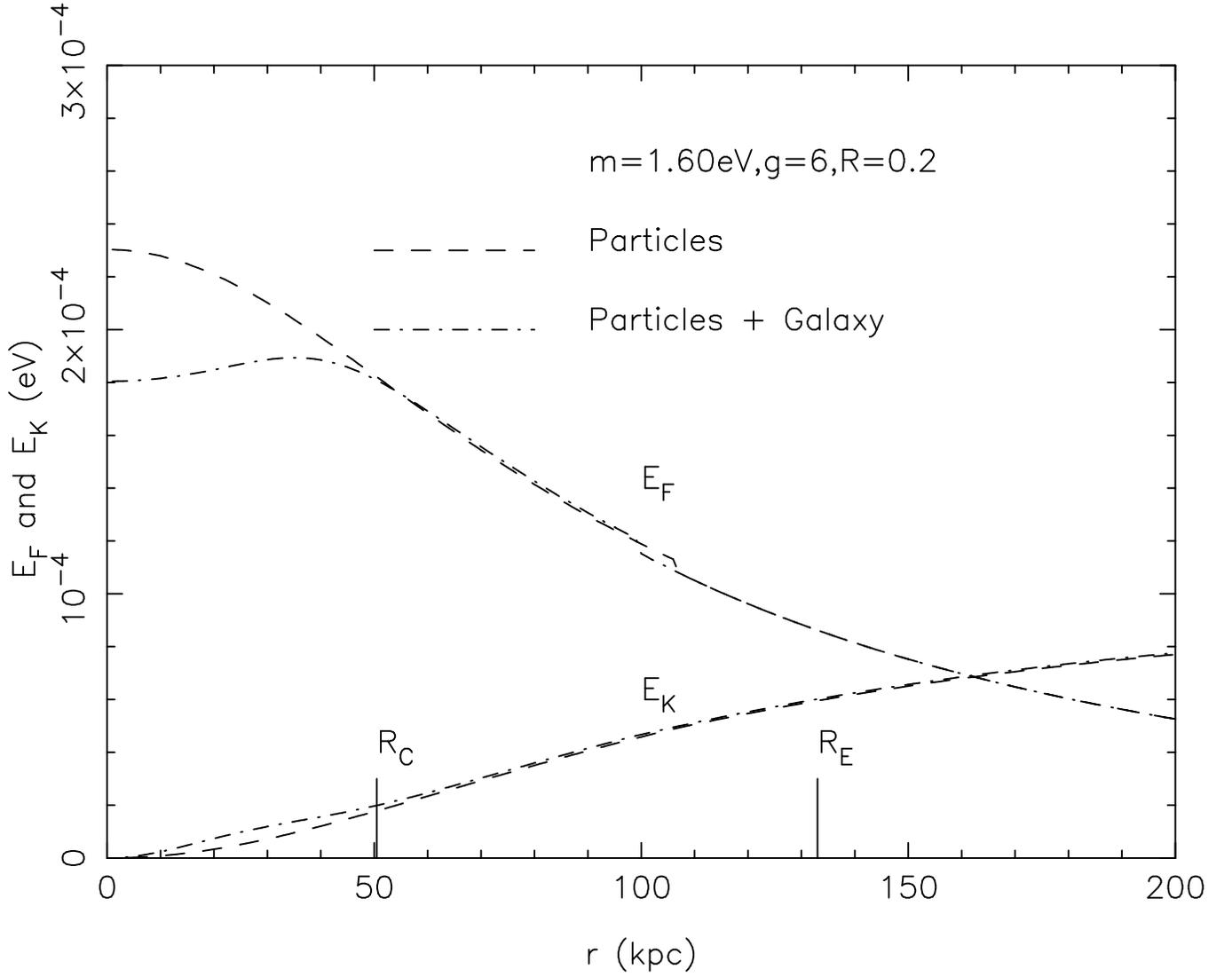}
\caption{Fermi levels and classical kinetic energies of 
a neutrino for the model without and with the toy galaxy. 
The model parameters are
$m$ = 1.6 eV, $g$=6, and $\rho({\rm B})/\rho({\rm DM}) = 0.2$.
There is a dip of the Fermi level for the model with the galaxy
indicating the local softening of the EOS due to the gravity of the
galaxy.  These energy levels are much higher than
those of unbound relic particles of $10^{-7}$ eV. \label{neu.EfEk}
}
\end{figure}

It is typical to have a small fraction ($<$2\%) of deficiency in the model 3D encircled
mass inside $r_E$ (1.06$\times10^{14}M_\odot$). A brightest cluster galaxy at
the center of a cluster, typically has a stellar mass in excess of $10^{12}M_\odot$
and a 2\% deficiency may be accounted for by a single massive galaxy at the center.
Since the mass profile obtained by \citet{TB05B} includes the contribution of galaxies,
we should at least examine the possible effect to our models dealing solely with particles.
Our toy galaxy has a total mass of  $2\times10^{12}M_\odot$ and a Gaussian volume profile
with the scale length of 20 kpc. We did not perform any fitting or adjustment of the galaxy
model parameters, except that the total mass corresponds to the 3D encircled mass deficiency 
with respect to the observed value. It turned out that this toy galaxy had no
effect on the particle mass, $m$, although the EHSE was solved from the origin as usual.
On the other hand, as can be seen in the volume- and column densities in Figs.\ref{neu.rho} and \ref{neu.Sigma},
the model galaxy remedies the local mass deficiency problem at the center. In Fig.\ref{neu.EfEk},
the Fermi levels $E_F$ and classical kinetic energies $E_K$ are plotted for the models without
and with the toy galaxy. The toy galaxy acts as an external gravity source, whose effect
on the 3D encircled mass is largest near the center. The gravity of the galaxy enhances
the classical kinetic energy of a particle and dissolves the degeneracy. This is the reason
why there is a dip of the Fermi level in the model with the galaxy. 
In other words, the EOS is softened locally by the stellar mass of the galaxy.
It should be
emphasized that this is true only if the galaxy is given as an external mass source
unaffected by the dark matter distribution. Also this effect is pronounced only near the
origin  of spherical symmetry, where the fractional contribution of the galaxy to the 3D encircled
mass is the greatest.

\section{Discussion}

\subsection{Relic Neutrinos and Bound Neutrinos in the Cluster}

One may wonder how fermions at the cluster center could be degenerate.
Here we first point out the fact that even relic neutrinos are moderately degenerate,
which used to be well known \citep{Weinberg1962}, but appears to be forgotten lately.
There is a clear distinction between the black body of fermions and that of bosons.
While the range of values of the Fermi distribution function 
for one-particle state is
between 0 and 1, that of the Bose distribution function 
is between 0 and $\infty$. So the fermion black body
is partially degenerate, while the boson black body is not.
Here we calculate the degree of degeneracy for relic neutrinos immediately after
decoupling and that after they become nonrelativistic, under the assumption of
adiabatic expansion.

The current cosmology assumes that there are the equal numbers of neutrinos and anti-neutrinos
for each species. So the present number density of each quantum statistically independent species,
$n(0)$, is
112.6/2 = 56.3 cm$^{-3}$. At redshift $z$, the number density $n(z)$ is given by

\begin{equation}
n(z) = 56.3 (1+z)^3 \hskip 2cm ({\rm cm}^{-3}).  
\end{equation}

Immediately after neutrino decoupling, 
when neutrinos were extremely relativistic, the Fermi level, $E_f$ is given by

\begin{equation}
E_f = c \left[6 \pi^2 \hbar^3 n(z)\right]^{1/3} = 2.94 \times 10^{-4} (1+z)\hskip 2cm ({\rm eV}). 
\end{equation}

On the other hand, the neutrino temperature then was

\begin{equation}
T_\nu = 1.95 (1+z)\hskip 2cm ({\rm K}). 
\end{equation}

Therefore the degree of degeneracy is given by

\begin{equation}
\frac{E_f}{k_B T_\nu} = 1.75, 
\end{equation}

which indicates moderate degeneracy. Under adiabatic expansion, this degree of degeneracy is unlikely
to change, while neutrinos are relativistic. There is a transition from the relativistic to nonrelativistic
regime when $k_B T = m c^2$.

Below

\begin{equation}
T_\nu = T_{tr} = \frac{m c^2}{k_B} = 1.16 m \times 10^4  \hskip 2cm ({\rm K}), 
\end{equation}

or at $z_{tr} \approx 6 m \times 10^3$, where $m$ is units of eV, 
neutrinos become nonrelativistic. Then the expressions for both Fermi level
and neutrino temperature change, but they have the same dependence on $(1+z)$.
As long as local gravity is negligible, nonrelativistic adiabatic expansion is assumed below $T_{tr}$,

\begin{equation}
E_f = \frac{\hbar^2}{2m} \left[6 \pi^2 n(z)\right]^{2/3} =  4.3\times10^{-8} m^{-1} (1+z)^2\hskip 2cm ({\rm eV}), 
\end{equation}

and 

\begin{equation}
T_\nu = T_{tr} \left(\frac{1+z}{1+z_{tr}}\right)^2 =  3.3\times10^{-4} m^{-1} (1+z)^2\hskip 2cm ({\rm K}),
\label{Tnonrela}
\end{equation}

or 

\begin{equation}
k_B T_\nu =  2.8\times10^{-8} m^{-1} (1+z)^2\hskip 2cm ({\rm eV}).
\label{KTnonrela}
\end{equation}

So the degree of degeneracy is 

\begin{equation}
\frac{E_f}{k_B T_\nu} = 1.53.
\end{equation}

Note that there is an offset of the zero point between the relativistic and nonrelativistic Fermi levels
given by $E_f({\rm relativistic}) = E_f({\rm nonrelativistic}) + m c^2$, where the offset is 
negligible when particles are extremely relativistic. 
Again in the nonrelativistic regime, relic neutrinos are expected to be moderately degenerate.
Due to the change of dispersion relation of individual particles caused by 
the relativistic-to-nonrelativistic transition, 
the degree of degeneracy changes slightly, but this effect appears
minor.

Next we compare the kinetic energy level of bound neutrinos in the cluster and
that of unbound neutrinos  to see if relic neutrinos are
cold enough so that they can fall to the cluster.
The Fermi level and gravitationally induced kinetic energy
at the cluster core are of $10^{-4}$ eV or above as seen in Fig.\ref{neu.EfEk}.
On the other hand,
unbound relic neutrinos which undergo adiabatic expansion have the temperature
given by eq(\ref{Tnonrela}).
For A1689 at $z=0.18$ and for $m=1.6$ eV, the kinetic energy of unbound relics will be $3.65\times 10^{-8}$ eV,
which is well below the energy level of the particles in the cluster core. So it is possible
that the relic neutrinos can fall into the cluster core.

The cluster also has hot plasma, which may affect the fermion degeneracy. Here we briefly argue that
the effect of hot plasma will be minor. 
First of all, hot plasma and neutrinos (or weakly interacting fermions in general) 
interact only through gravity and the plasma temperature
has nothing to do with the velocity dispersion of neutrinos, which is determined by
the encircles mass at a given $r$. 
What matters is the plasma's contribution to the encircled mass (or encircled gravitational energy if
the plasma is relativistic).
The ratio of gas mass and total mass has been estimated to
be about 7\% by X-ray observations \citep{Anderson04}. They interact with fermions only through
Newtonian gravity, because the plasma is not hot enough to be relativistic or the kinetic energy
of a plasma particle
is negligible compared to the rest-mass energy.
Therefore unless
the plasma particles have exceptionally high concentration near the 3D center of the cluster,
their effect on the encircled mass should be minor compared to that by the dark matter particles.
However, an extreme high concentration is not expected unless the plasma temperature
profile has a singularity near the center. Since the X-ray emission is integrated along the light of sight,
X-ray observations alone will  not be capable of pin-pointing the plasma temperature near the 3D center of the cluster.

\subsection{Direction of the Future Study}

This degenerate fermion/neutrino hypothesis should be tested
observationally with future-coming density profiles of other clusters obtained
by gravitational lensing.
Those profiles should be modeled by
the fixed set of particle properties, $(m,g)$, and by a varied set 
of the central density, $\rho(0)$. Modeling multiple cluster profiles may further constrain 
the particle properties.

There may be a room for improvement in the phenomenological equation of state (PEOS).
The PEOS used in this paper, eq(\ref{peos}), overestimates the degeneracy pressure, $p_D$,
since $p_D$ here is the zero-temperature degeneracy pressure for a given number density, $n$,
regardless of the temperature of fermions or regardless of the degree of degeneracy.
For the same observed density profile, the best-fit model with
a more realistic PEOS will require a higher number density, $n$, or 
a lower particle mass. 
Another possible improvement is that without fixing the density ratio of total matter and light fermions,
the densities of light fermions and collisionless cold dark matter particles can be treated
separately in terms of the equation of hydrostatic equilibrium for the former 
and the Jean equation for the latter, while they are coupled through the
total encircled mass and the common velocity dispersion at a given $r$.

One cosmological subject related particularity to the degeneracy of relic massive neutrinos, is
its potential effect on ``free streaming''. Relic neutrinos are hot dark matter, which were relativistic,
when they decoupled. From the picture of completely collisionless particles, it is
considered that small-scale
fluctuations in the primordial neutrino density fluctuation spectrum 
should have been wiped out. However, the partial degeneracy of relic neutrinos implies that
not all of them could have behaved as free particles.
It is conceivable that the effect of Fermi-Dirac degeneracy is to suppress the ``free streaming'' or to preserve the initial
fluctuation spectrum.
Depending on the degree of this ``free-streaming suppression''
of neutrinos or fermionic hot dark matter in general, our expectation on the initial density fluctuation
spectrum  may need to be revised. We plan to investigate this potential effect quantitatively
in the near future.

\section{Concluding Remark}
 
In this paper, 
we have shown that the observed mass profile at the center of the cluster of galaxies,
A1689, is reproduced by models assuming 
the presence of degenerate fermionic dark matter.
In the case that previously unknown fermions with spin 1/2 dominate
the dark matter, the acceptable particle mass range is between 2 and 4 eV.
In the case that the dark matter consists of the mixture of 
the degenerate relic neutrinos and classical collisionless cold dark matter particles,
the mass range of neutrinos is between 1 and 2 eV, if the ratio of the two kinds
of dark matter particles is fixed to its cosmic value.
Both the pure fermionic dark matter models and neutrino-CDM-mixture models reproduce the observations
equally well.







\acknowledgments
We thank the anonymous referee for useful comments.
This study was motivated by a very interesting talk given by Tom
Broadhurst in 2004 at NAOJ, Mitaka, Japan.

\clearpage

\begin{table}
\begin{center}
\caption{Solutions for a Single Species of Spin 1/2 Fermions as the Dominant Dark Matter\label{fermion.table}}
\vskip 3mm
\begin{tabular}{ccccc}
\tableline\tableline
$\rho({\rm B})/\rho({\rm DM})$ & Particle Type & $g$ & $(m,r_t)$  & Profile\\
                               &               &     & eV,kpc     &        \\
\tableline
0.2   & Majorana (p) & 1 & (3.9,104) $\sim$ (4.3,73) & B \\
      & Dirac (p) or Majorana (p+$\bar{\rm p}$) & 2  & (3.3,93) $\sim$ (3.6,70) & I \\
      & Dirac (p+$\bar{\rm p}$) & 4 & (2.8,100) $\sim$ (3.1,73)  & A \\   
\tableline
0.5   & Majorana (p) & 1 & (3.4,100) $\sim$ (3.7,73) & A \\
      & Dirac (p) or Majorana (p+$\bar{\rm p}$) & 2  & (2.9,106) $\sim$ (3.1,73) & I \\
      & Dirac (p+$\bar{\rm p}$) & 4 & (2.4,100) $\sim$ (2.6,73) & A \\   
 \tableline
\end{tabular}
\end{center}
\end{table}

\clearpage

\begin{table}
\begin{center}
\caption{Solutions for the Mixture of Neutrinos and Nondegenerate Cold Dark Matter\label{neutrino.table}}
\vskip 3mm
\begin{tabular}{cccc}
\tableline\tableline
$\rho({\rm B})/\rho({\rm DM})$ & Particle Type & $g$  & $(m,r_t)$ \\
                               &               &      &  eV,kpc   \\
\tableline
0.2   & $\nu + \bar{\nu}$ & 6 & (1.6,106) $\sim$ (2.1,73) \\
\tableline
0.5   & $\nu + \bar{\nu}$ & 6 & (1.1,106) $\sim$ (2.4,73) \\
\tableline
\end{tabular}
\end{center}
\end{table}




\clearpage

\clearpage

\clearpage

\clearpage

\clearpage

\clearpage

\clearpage

\clearpage

\clearpage

\end{document}